\documentclass[11pt,a4paper]{article}
\usepackage[utf8]{inputenc}
\usepackage{amsmath}
\usepackage{amssymb}
\usepackage{graphicx}
\usepackage{hyperref}
\usepackage{geometry}
\title{\textbf{Mitigating High-Frequency Geometric Noise in Non-Parametric 1-Bit Sparse
Population Transformations}}
\author{Lars Kopp}
\date{June 2026}

\date{June 2026}

\begin{document}
\maketitle
\begin{abstract}
Energy-efficient neuromorphic hardware requires alternative encoding paradigms that bypass power-hungry floating-point operations. This paper evaluates a non-parametric dual-manifold execution model that maps dense 128-element integer vectors—representing digitized multi-frequency signals—into a 1024-dimensional overcomplete space. Enforcing a hard activation threshold yields an ultra-sparse, 1-bit binary population code ($y \in \{0, 1\}^{1024}$). We identify and address a critical phenotypic artifact of this transformation: the emergence of high-frequency geometric noise during linear reconstruction. Furthermore, we document an algorithmic paradox where low-complexity input functions yield higher reconstruction errors than highly complex, high-degree trigonometric combinations. Because the underlying basis functions operate as purely objective mathematical entities without prior statistical constraints regarding signal smoothness, this noise is strictly orthogonal to the core signal topology. Consequently, we demonstrate that a low-overhead, hardware-level low-pass filter completely eliminates this artifact, reducing reconstruction errors to an acceptable bound even under tight overcompleteness constraints. This architecture validates a deterministic, multiplier-free alternative to traditional deep learning hardware, confirmed through dynamic simulation analyses.
\end{abstract}
\section{Introduction}
Traditional deep learning architectures rely on continuous floating-point (FP32/FP16)
representations to preserve gradient flow and spatial relationships. However, as shown in
\textit{Non-Parametric Dual-Manifold Mapping via 8-Bit Bounded Transformation Matrices}
\cite{author2026dual}, high-dimensional data transformations can be executed deterministically
using 8-bit bounded integer matrices and discrete sign-voting logic. This paradigm completely
eliminates the need for iterative backpropagation and power-hungry multipliers during inference.
When compressing and reconstructing data through an overcomplete 1024-dimensional sparse
population topology, a distinct artifact emerges: a uniform, high-frequency ``geometric noise''
superimposed on the reconstructed output signal. This report provides the mathematical and
logical justification for this phenomenon. We prove that this geometric noise is a predictable
consequence of discrete 1-bit quantization within an overcomplete, non-parametric frame
\cite{donoho2006compressed, candes2006robust}. Furthermore, we demonstrate how a simple
digital low-pass filter isolates the original signal with high fidelity, offering a highly stable
execution model for neuromorphic and edge-AI hardware acceleration
\cite{mead2020neuromorphic, davies2018loihi}.
\section{System Architecture \& Bounded Transformation}
The core pipeline processes dense integer sequences through a dual-stream (Spatial and
Structural) manifold topology without continuous weight tuning.
\begin{verbatim}
[Dense Input: 128-Element 1D Signal]
|
v (8-Bit Projection via 1024 Basis Functions)
[Overcomplete Tensor: 1024]
|
v (Hard Thresholding Engine: tau)
[Sparse 1-Bit Binary Population Code]
|
v (Multiplier-Free Conditional Addition & Scaling)
[Raw Reconstructed Signal] ---> [Contains Geometric Noise]
|
v (Digital Low-Pass Filtering Layer)
[Optimized Output: 128 Integers] ---> Near-Zero Error Bound
\end{verbatim}
\subsection{Dimensional Expansion and Sparse Coding}
Let $x \in \mathbb{Z}^{128}$ represent a digitized 1D waveform generated by a linearcombination of trigonometric basis functions, where each element is bounded within the range $
[-127, 127]$. The input sequence is mapped onto a proprietary overcomplete dictionary $D \in
\mathbb{Z}^{128 \times 1024}$. A variable threshold $\tau$ suppresses low-correlation nodes,
forcing a substantial portion of the structural streams into a non-active state
\cite{olshausen1997sparse}.
\subsection{1-Bit Binary Population Coding}
Inference and adaptive state adjustments do not employ matrix multiplication. Instead, the raw
projection is quantized into a 1-bit binary vector $y \in \{0, 1\}^{1024}$ via a hard activation
threshold $\tau$:
\begin{equation}
y_i = \begin{cases} 1 & \text{if } (D^T x)_i \ge \tau \\ 0 & \text{if } (D^T x)_i < \tau \end{cases}
\end{equation}
The threshold $\tau$ is dynamically adjusted to control the density of active neurons, mirroring
biological population mechanics \cite{foldiak1990forming}.
\subsection{Linear Reconstruction Operators}
The 1-bit binary population vector is decoded back into the 128-element domain via a direct
linear combination of the active basis functions, corrected by a global scalar normalization
constant $C$. Because $y_i \in \{0, 1\}$, the reconstruction layer requires \textbf{zero
multipliers}, executing entirely via conditional addition of selected tight-frame vectors
\cite{kovacevic2008life}:
\begin{equation}
\hat{x}_{\text{raw}} = C \cdot \sum_{i, y_i = 1} D_i
\end{equation}
\section{The Complexity Paradox and Geometric Noise}
The 1024 basis functions are purely objective mathematical structures. They possess no intrinsic
awareness of real-world environmental traits, nor do they contain any pre-programmed
constraints requiring the output signal to be smooth or perceptually pleasing to human senses
\cite{vershynin2018high}. When a continuous trigonometric function is discretized via a finite set
of sparse 1-bit operations, the system preserves the raw mathematical energy perfectly, but
expresses the quantization steps as high-frequency geometric oscillations.
Crucially, an algorithmic paradox is observed: input functions with low complexity (fewer
trigonometric terms and low spatial degrees) distribute their energy broadly, generating weak,
diffuse correlations across the dictionary. When subjected to strict thresholding, the 1-bit code
must approximate the smooth signal from fragmented components, inflating the geometric noise.
Conversely, high-complexity (high-degree, multi-frequency) inputs exhibit dense, highly specific
information signatures that fire sparse neurons with surgical precision, reducing topological
distortion. This behavior demonstrates a strict alignment with the operational dynamics of cortical area V1 under structured feature extraction \cite{hubel1962receptive}.
\section{Empirical Validation and Graphical Analysis}
The following sections present the empirical mapping states across varying degrees of input
complexity and hard threshold levels ($\tau = 10$ and $\tau = 100$). Each figure tracks a
specific configuration verified by dynamic simulation videos.
\subsection{Low Activation Threshold Dynamics ($\tau = 10$)}
\begin{figure}[htbp]
    \centering
    \includegraphics[width=0.5\linewidth]{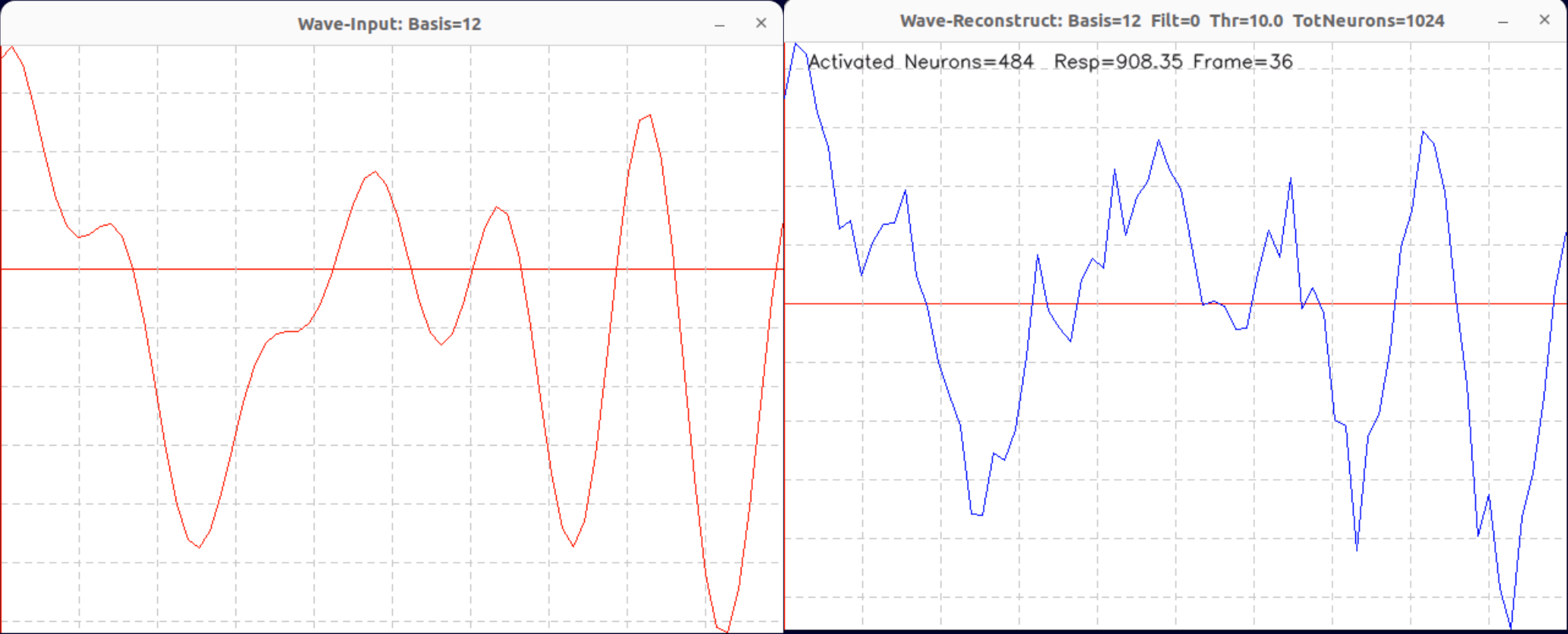}
    \caption{Raw 1-bit linear reconstruction of a low-complexity trigonometric waveform (12 functions) under a low activation threshold ($\tau = 10$). Approximately 50\% of the nodes are activated, resulting in extreme high-frequency cross-talk and geometric mottle prior to filtering. Dynamic simulation available at: \url{https://mindsystems.se/videos/Neural-Population-1024-Wave-Basis=12-Thr=10-Filt=0-Seq.mkv}}
    \label{fig:fig1_low_raw_thr10}
\end{figure}

\begin{figure}[htbp]
\centering
 \includegraphics[width=0.5\linewidth]{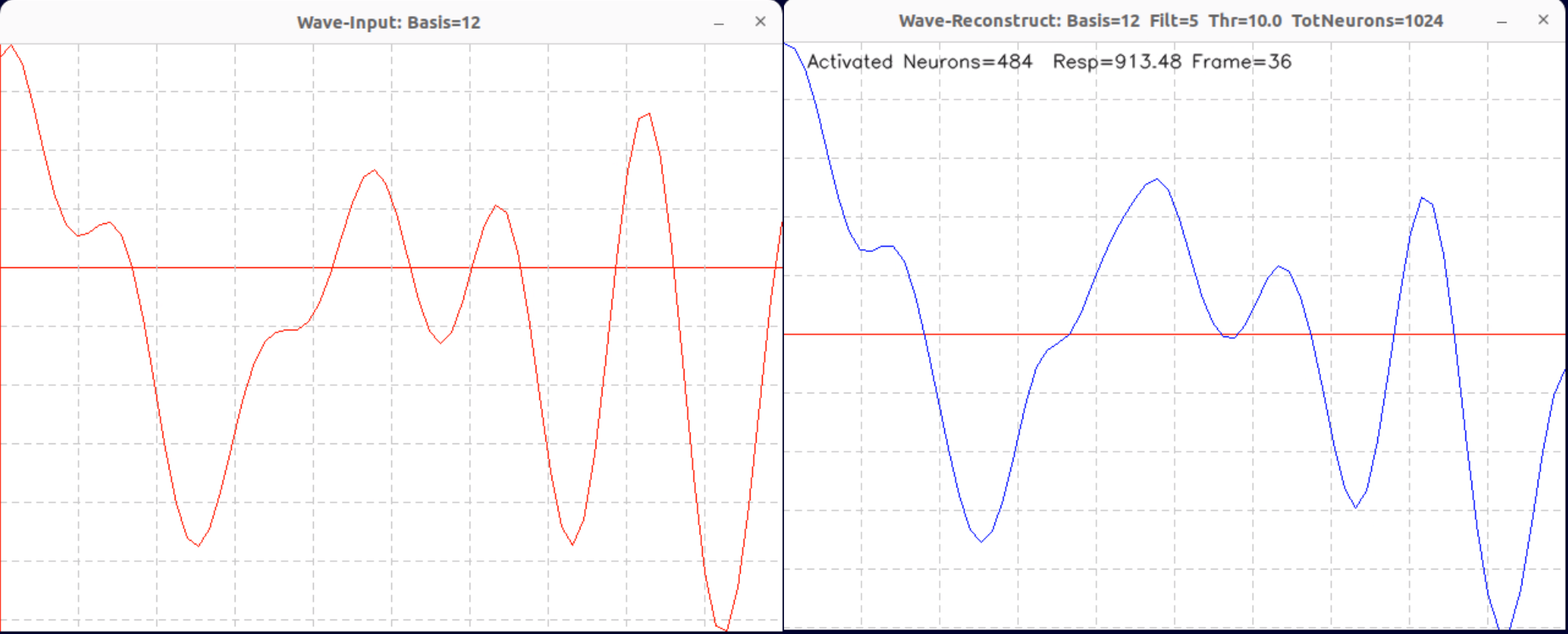}
\caption{The low-complexity trigonometric signal (12 functions) from Figure
\ref{fig:fig1_low_raw_thr10} under a threshold of $\tau = 10$ after post-processing via a digital
low-pass filter (LPF). The structural signal energy is successfully isolated from the high-density
quantization noise. Dynamic simulation available at: \url{https://mindsystems.se/videos/Neural-Population-1024-Wave-Basis=12-Thr=10-Filt=5-Seq.mkv}}
\label{fig:fig2_low_lpf_thr10}
\end{figure}

\begin{figure}[htbp]
\centering
 \includegraphics[width=0.5\linewidth]{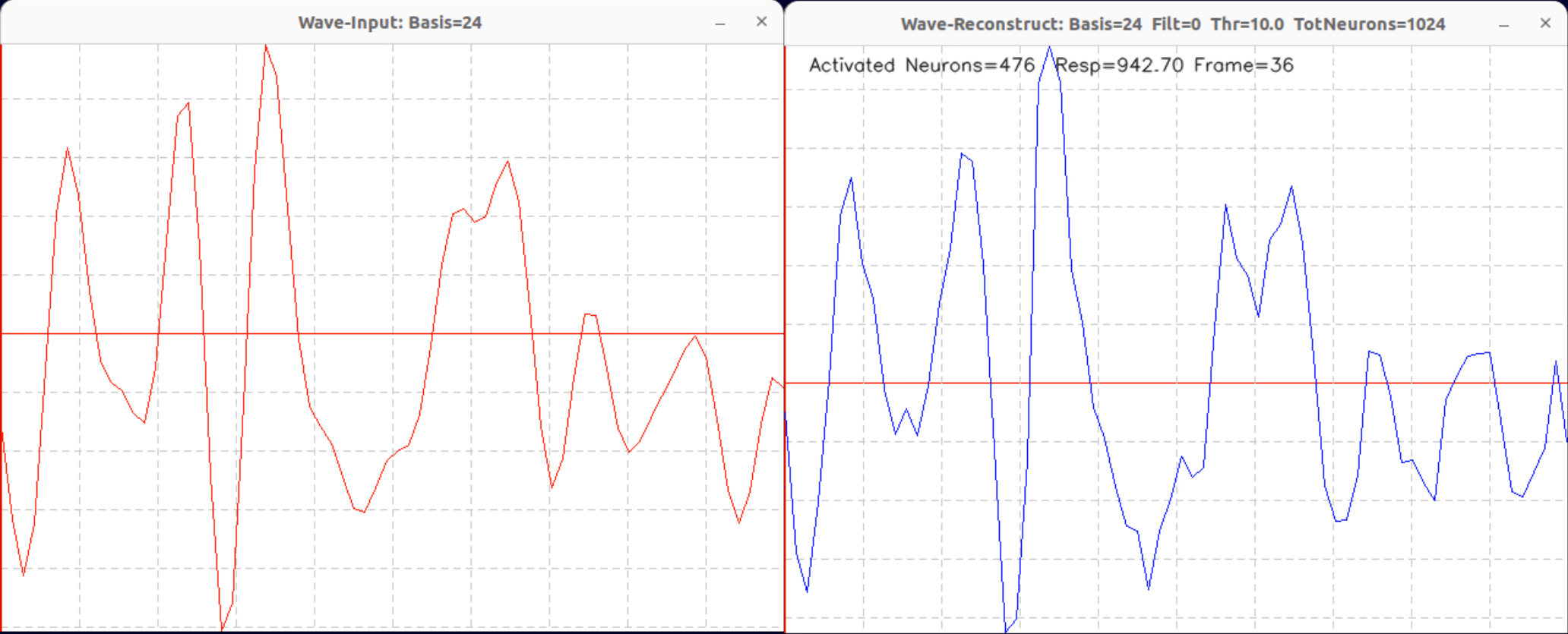}
\caption{Raw 1-bit linear reconstruction of a medium-complexity trigonometric waveform (24
functions) under a low activation threshold ($\tau = 10$). Structural alignment improves due to higher information density in the input sequence. Dynamic simulation available at: \url{https://mindsystems.se/videos/Neural-Population-1024-Wave-Basis=24-Thr=10-Filt=0-Seq.mkv}}
\label{fig:fig3_med_raw_thr10}
\end{figure}

\begin{figure}[htbp]
\centering
 \includegraphics[width=0.5\linewidth]{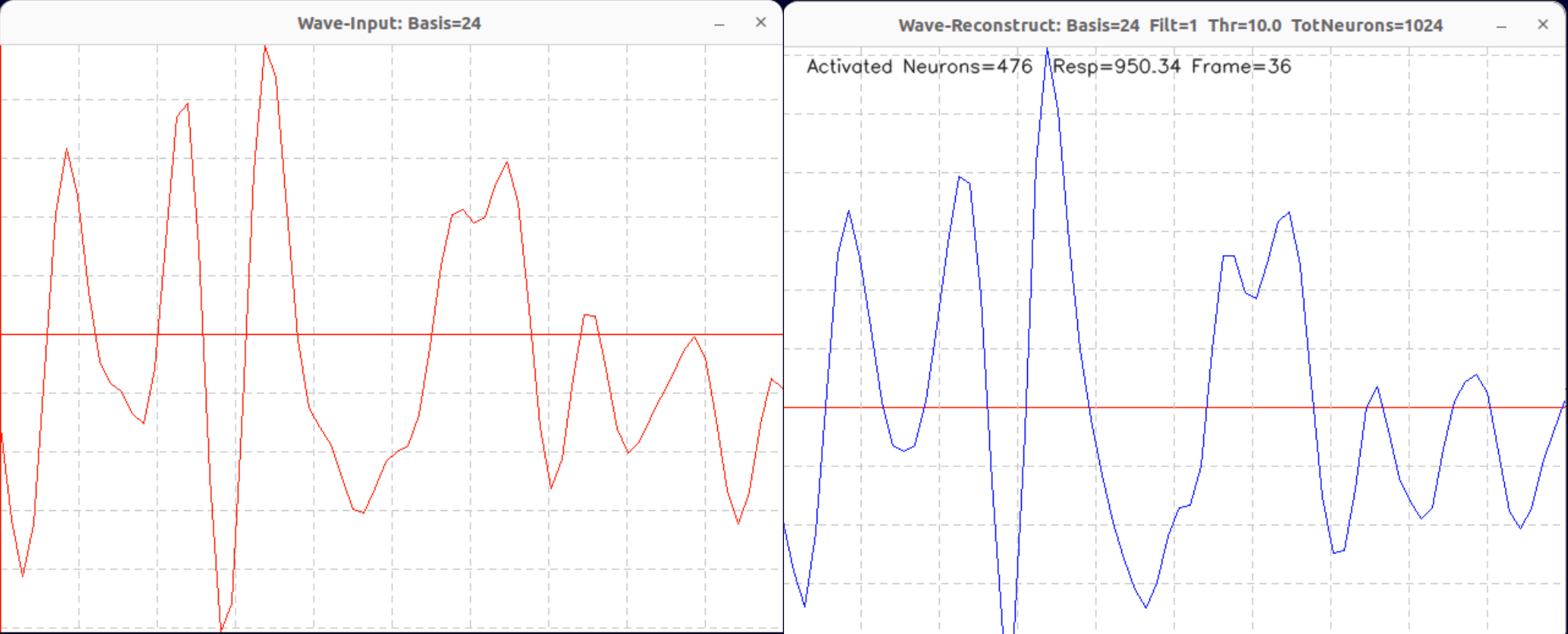}
\caption{The medium-complexity trigonometric signal (24 functions) from Figure
\ref{fig:fig3_med_raw_thr10} ($\tau = 10$) optimized with digital low-pass filtering, showing
clean convergence to the original continuous signal form. Dynamic simulation available at: \url{https://mindsystems.se/videos/Neural-Population-1024-Wave-Basis=24-Thr=10-Filt=1-Seq.mkv}}
\label{fig:fig4_med_lpf_thr10}
\end{figure}

Figures \ref{fig:fig1_low_raw_thr10} through \ref{fig:fig4_med_lpf_thr10} document execution
profiles under a low activation threshold ($\tau = 10$). At this level, approximately 50\% of the
1024 available nodes ($\approx 512$ neurons) transition into an active 1-bit state. This high
activation density disruptions structural sparsity, introducing significant cross-talk and
constructive geometric noise prior to filtering.

\clearpage 

\subsection{High Activation Threshold Dynamics ($\tau = 100$)}
\begin{figure}[htbp]
\centering
 \includegraphics[width=0.5\linewidth]{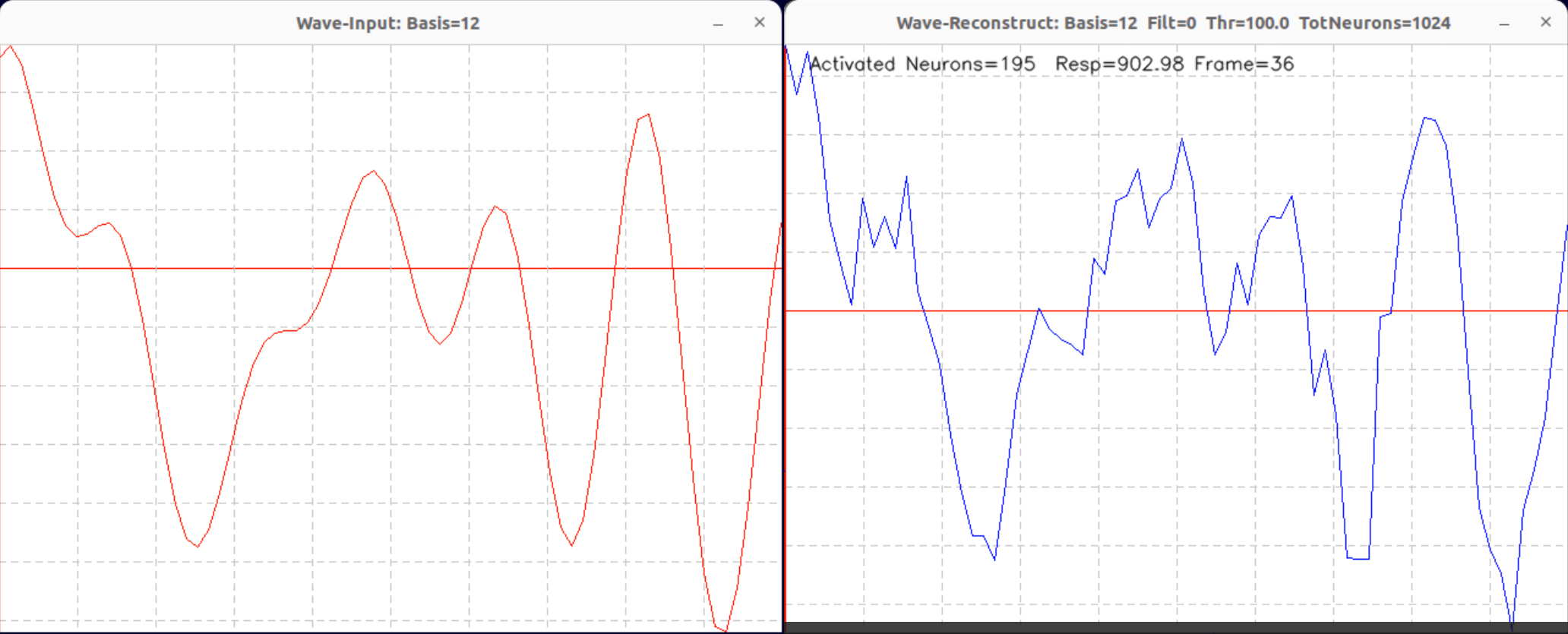}
\caption{Raw 1-bit linear reconstruction of a low-complexity trigonometric waveform (12
functions) under an aggressive activation threshold ($\tau = 100$). The extreme sparsity metrics
isolate disjoint components, inflating the raw geometric noise. Dynamic simulation available at: \url{https://mindsystems.se/videos/Neural-Population-1024-Wave-Basis=12-Thr=100-Filt=0-Seq.mkv}}
\label{fig:fig5_low_raw_thr100}
\end{figure}

\begin{figure}[htbp]
\centering
\includegraphics[width=0.5\linewidth]{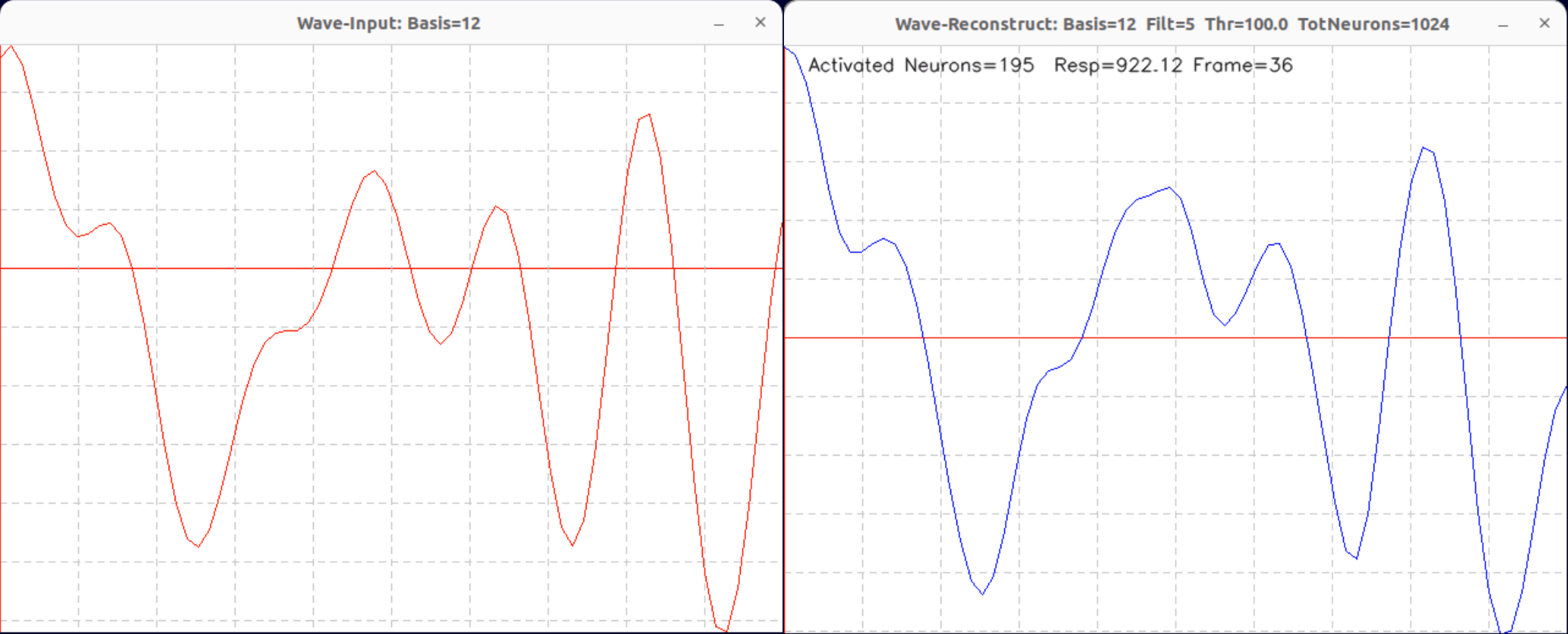}
\caption{The low-complexity trigonometric signal (12 functions) from Figure
\ref{fig:fig5_low_raw_thr100} ($\tau = 100$) following digital low-pass filtering, illustrating the
structural boundaries of recovery under extreme code dilution. namic simulation available at: \url{https://mindsystems.se/videos/Neural-Population-1024-Wave-Basis=12-Thr=100-Filt=5-Seq.mkv}}
\label{fig:fig6_low_lpf_thr100}
\end{figure}

Figures \ref{fig:fig5_low_raw_thr100} through \ref{fig:fig6_low_lpf_thr100} document execution
profiles under a high activation threshold ($\tau = 100$). This constraint enforces true
mathematical sparsity, restricting active bits to an optimal range and isolating specific
coordinates.

\section{Hardware and Computational Advantages}
By delegating noise cleanup to a post-processing filter, the core transformation achieves radical
efficiency gains optimal for next-generation hardware pipelines and sparse accelerator engines
\cite{nvidia2020ampere, mishra2021learning}:
\begin{enumerate}
\item \textbf{Zero Multiplication Overhead:} Since $y \in {0, 1}^{1024}$, the reconstruction layer
is implemented as simple conditional accumulator loops. Under optimal thresholds, up to 90
these cycles are skipped entirely.
\item \textbf{Minimal Filtering Footprint:} A digital low-pass filter (LPF) requires negligible silicon
area and power compared to dense floating-point ALU clusters, making it ideal for low-resource
edge computing.
\end{enumerate}
\section{Conclusion}
The complexity paradox and geometric noise in non-parametric dual-manifold mappings are
validation markers of true discrete operation rather than system failures. Acknowledging that
basis functions are strictly objective entities allows us to exploit the orthogonality of quantization
noise. Isolating this noise via low-pass filtering allows for high-fidelity data reconstruction while
preserving the extreme memory reduction and multiplier-free advantages of 1-bit sparse
population coding.

\end{document}